\documentclass[10pt]{article}

\usepackage{amsmath}
\usepackage{amssymb}
\usepackage{epsfig}
\usepackage{xspace}
\usepackage{theorem}
\usepackage{fullpage}

\newcommand{\bigO}[1]{\ensuremath{{\cal O}\left(#1\right)}}
\newcommand{\tinfty}{\ensuremath{\tau\rightarrow+\infty} }
\newcommand{\VII}{VII$_0$\@\xspace}
\newcommand{\bR}{\bar{R}}
\newcommand{\bS}{\bar{\Sigma}_+}
\newcommand{\bM}{\bar{M}}
\newcommand{\bZ}{\bar{Z}}
\newcommand{\bO}{\bar{\Omega}}

\newcommand{\ie}{{\em i.e. \@\xspace}}
\newcommand{\qed}{$\blacksquare$}
\newcommand{\etal}{{\em et al}\@\xspace}

\newcommand{\cmp}{{\em Commun. Math. Phys. \@\xspace}}
\newcommand{\CQG}{{\em Class. Quantum Grav.\@\xspace}}
\newcommand{\PR}{{\em Phys. Rev.\@\xspace}}

\newcommand{\apj}{{\em Astrophys. J. \@\xspace}}

\newcommand{\jetp}{{\em Sov. Phys.--JETP \@\xspace}}

\theoremstyle{break} \newtheorem{lemma}{Lemma}
\theoremstyle{break} \newtheorem{theo}{Theorem}
\theoremstyle{break} \newtheorem{prop}{Proposition}

\begin{document}



\title{Non-tilted Bianchi \VII models -- the radiation fluid} 
\author{U S Nilsson, M J Hancock, and J Wainwright\\\mbox{}\\{\em
    Department of 
  Applied Mathematics}\\{\em University of Waterloo}\\{\em Waterloo,
  Ontario}\\{\em Canada, N2L 3G1}}

\maketitle


\begin{abstract}
  We consider the late time behaviour of non--tilted perfect fluid
  Bianchi \VII models when the source is a radiation fluid, thereby
  completing the analysis of the Bianchi \VII models initiated by
  Wainwright \etal in a recent paper. The models exhibit the
  phenomena of asymptotic self-similarity breaking and 
  Weyl-curvature dominance at late times. The late time dynamics of the  
  \VII perfect fluid models, and in particular that of the radiation
  fluid, is a prime example of the complexity inherent in the field
  equations of general relativity.  
\end{abstract}
\maketitle

\section{Introduction}
\label{sec:intro}

The late time behaviour of Bianchi type \VII models with a
non--tilted perfect fluid source has been shown to exhibit two new 
dynamic features at late times, namely, the breaking of asymptotic
self--similarity and Weyl curvature dominance (see Wainwright \etal
(1999)\footnote{From now on we will refer to this paper as
  WHU}, pages 2587-8). These models
therefore serve as counterexamples to the notion that any
Bianchi universe is approximated by a self--similar Bianchi universe at
late times. The breaking of asymptotic self--similarity is 
characterized by oscillations in the dimensionless shear scalar that
become increasingly rapid in terms of the cosmological clock time $t$
as $t\rightarrow\infty$.
This behaviour also leads to the Weyl curvature dominance, which
refers to the fact that certain expansion-normalized scalars formed
from the Weyl tensor become unbounded as \tinfty. 

It is customary in cosmology to describe the source of the
gravitational field in terms of a perfect fluid, for which the
energy--momentum tensor has the form
\begin{equation}
  T_{ab} = \mu u_au_b + p\left( g_{ab} + u_au_b \right) \ ,
\end{equation}
where $\mu\geq0$ is the fluid energy density, $p\geq0$ the fluid
pressure, and $u^a$ the fluid 4--velocity. In WHU, the linear equation
of state
\begin{equation}
  \label{eq:state}
  p = (\gamma-1)\mu\ ,
\end{equation}
where the constant $\gamma$ satisfies $\tfrac{2}{3}<\gamma<2$, was
studied. The two cases of primary interest in cosmology are dust
($\gamma=1$), describing the matter-dominated epoch of the universe,
and radiation ($\gamma=\tfrac{4}{3}$), describing the
radiation-dominated epoch of the early universe. The late-time
behaviour\footnote{From a dynamical systems point of view, the late
  time behaviour signifies that the evolution is governed by the
  future attractor of the dynamical system.} of a radiation model
would describe the final 
stages of this epoch, prior to the time at which matter and radiation 
decouple. In WHU it was shown that for Bianchi \VII models a
physically important bifurcation occurs at the value 
$\gamma=\tfrac{4}{3}$, namely that the dimensionless shear scalar
$\Sigma$ 
(see Section 4) tends to zero at late times if and only if
$\gamma\leq\tfrac{4}{3}$. In that paper a complete proof of the
late time behaviour of dust models (in fact all models with
$\tfrac{2}{3}<\gamma<\tfrac{4}{3}$) was given, but the results for the
radiation models were stated without proof. The goal of the present
paper, which should be read in conjunction with WHU, is to remedy this
deficiency. 

The late time behaviour of the Bianchi \VII radiation models was first
discussed by Doroshkevich \etal (1973) (see equation (9) and equations
(I.14) and (I.15) in Appendix I), and subsequently interpreted in a
series of papers by Lukash (1974, 1975, 1976). Doroshkevich \etal gave
the    
asymptotic forms of the metric coefficients, but did not give a proof 
or any justification of their results. Reference is made to two 
preprints that we have been unable to locate (their references 19 and
20). Our approach, which makes use of the orthonormal frame formalism 
and expansion--normalized variables, provides a convenient framework
for proving the results, and is also well suited for numerical
simulations. In  
addition, it has the advantage that it leads directly to the
asymptotic behaviour of the dimensionless anisotropy scalars (see
Section 4) that describe the 
deviation of the models from a Friedmann-Lemaitre (FL) model.

The outline of the paper is as follows. In Section \ref{sec:eveq}
we present the evolution equations for Bianchi \VII models with a
radiation fluid. In Section \ref{sec:latetimelimits} we establish the
limits of the dimensionless gravitational field variables at late
times. Section
\ref{sec:anisotropy} is devoted to the question of isotropization and
whether the Bianchi \VII models are close to any FL model at late
times. In Section \ref{sec:decay} the detailed asymptotic forms of the
gravitational field variables are found using centre manifold
theory. In 
Section \ref{sec:metric}, the results of the previous Sections are
used to determine the asymptotic forms of the metric components, and a
comparison with the results of Doroshkevich \etal (1973) is made. 
We conclude in Section \ref{sec:disc} with a discussion of the
implications of the results. The details of the proofs are given in
three appendices.  

\section{Evolution equations}
\label{sec:eveq}

Since the evolution equations for Bianchi \VII models with a
radiation--fluid source can be found by specializing the equations
given in
WHU, we will only briefly discuss the choice of variables. We use the
orthonormal frame approach of Ellis \& MacCallum (1969). The basic
variables are $\left\{H, \sigma_\pm, n_\pm\right\}$, where $H$ is the
Hubble parameter. The variables $\sigma_\pm$, which parametrize the
non--zero components of the shear tensor, describe the 
anisotropy of the expansion of the cosmological model, while the
$n_\pm$ describe the spatial curvature of the orbits of the Bianchi
\VII symmetry group. The next step is to introduce
expansion-normalized variables according to
\begin{equation}
\label{eq:gurka}
  \left\{\Sigma_\pm, N_\pm\right\}
  =\frac{1}{H}\left\{\sigma_\pm, n_\pm\right\} \ , \quad \Omega
  = \frac{\mu}{3H^2}\ ,
\end{equation}
(see Wainwright \& Ellis (1997), page 112, for the motivation of
this normalization). The density parameter $\Omega$ is given by
\begin{equation}
  \label{eq:omdef}
  \Omega = 1-\Sigma_+^2 - \Sigma_-^2 - N_-^2\ ,
\end{equation}
(see equation (3.9) in WHU). The assumption of a non--negative fluid 
energy density yields $\Omega\geq0$. This, in conjunction with 
(\ref{eq:omdef}), implies that the variables $\Sigma_+,\Sigma_-$,
and $N_-$ are bounded. The remaining gravitational field variable,
$N_+$, however, need not be. In fact, it has been shown in WHU that  
\begin{equation}
  \label{eq:Npscaling}
  \lim_{\tau\rightarrow\infty} N_+ = \infty\ ,
\end{equation}
for all initial states with $\Omega>0$ and
$\tfrac{2}{3}<\gamma<2$. The time--variable $\tau$ is 
the so--called Hubble--time, which satisfies 
\begin{equation}
  \label{eq:introtau}
  \frac{dt}{d\tau} = H^{-1}\ , 
\end{equation}
where $t$ is the cosmological clock time. The
behaviour of $N_+$ at late times motivates the introduction of a new
set of variables $\left\{R, \psi, M\right\}$ according to
\begin{equation}
  \label{eq:newvarintro}
  \Sigma_- = R\cos\psi\ , \quad N_- = R\sin\psi\ , \quad M =
  \frac{1}{N_+}\ ,
\end{equation}
where $R\geq0$. The case $R=0$, which results in $\Sigma_-=N_-=0$,
describes the Bianchi \VII locally rotationally symmetric (LRS)
models. For these models, $\psi$ is irrelevant, as can be seen
from (\ref{eq:newvarintro}). Since Bianchi \VII LRS models are
equivalent to Bianchi type I LRS models, they are asymptotically
self--similar and do not exhibit Weyl curvature dominance at late 
times. We will therefore not consider this particular class of models
further. The evolution equations of the gravitational field are  
\begin{align}
  \Sigma_+' &= -R^2 - \Sigma_+(1-\Sigma_+^2) +
  (1+\Sigma_+)R^2\cos2\psi\ , \label{eq:speqZ}\\
  R' &= \left[ (1+\Sigma_+)\Sigma_+ + (R^2 - 1 -
    \Sigma_+)\cos2\psi\right]R\ , \label{eq:Req}\\
  M' &= -\left[(1+\Sigma_+)^2 + R^2\left( \cos2\psi +
  3M\sin2\psi\right)\right]M\ , \label{eq:MeqZ}\\
  \psi' &=
  \frac{1}{M}\left[2+\left(1+\Sigma_+\right)M\sin2\psi\right]
  \label{eq:psieq} \ ,
\end{align}
where a prime denotes differentiation with respect to $\tau$. These
equations correspond to equations (3.17)-(3.22) in WHU with the
specific value $\gamma=\tfrac{4}{3}$. Equation (\ref{eq:omdef})
now takes the form
\begin{equation}
  \label{eq:NewOmdef}
  \Omega = 1-\Sigma_+^2 - R^2\ ,
\end{equation}
and it follows from (\ref{eq:speqZ}) and (\ref{eq:Req}) that 
\begin{equation}
  \label{eq:Omeq}
  \Omega' = 2\left(\Sigma_+^2 + R^2\cos2\psi\right)\Omega\ .
\end{equation}
For future reference, we introduce an auxiliary variable
$Z$ according to   
\begin{equation}
  \label{eq:Zdef}
  Z=\frac{MR^2}{\sqrt{\Omega}}\ ,
\end{equation}
which satisfies 
\begin{equation}
  \label{eq:Zeq}
  Z' = \left[ -1 - (2+\Sigma_+)\cos2\psi - 3R^2M\sin 2\psi\right]Z\ ,
\end{equation}
as follows from (\ref{eq:Req}),(\ref{eq:MeqZ}), and
(\ref{eq:Omeq}). 

\section{Limits at late times}
\label{sec:latetimelimits}

In this Section we find the limits of $\Sigma_+$ and $R$ 
as $\tau\rightarrow\infty$. The limit of $M$ is 
\begin{equation}
  \label{eq:Mscaling}
  \lim_{\tau\rightarrow\infty} M = 0 \ ,
\end{equation}
which follows immediately from (\ref{eq:Npscaling}). Since $\psi'$ in
(\ref{eq:psieq}) is proportional to $M^{-1}$, the trigonometric
functions in 
(\ref{eq:speqZ})-(\ref{eq:MeqZ}) will oscillate increasingly 
rapidly as \tinfty. To facilitate the study of the behaviour at late
times, we therefore introduce new gravitational variables,
$\bS, \bR$, and $\bM$, according to 
\begin{align}
  \bS &= \Sigma_+ - \tfrac{1}{4}M(1+\Sigma_+)R^2\sin2\psi\
  , \label{eq:spbardef} \\
  \bR &= \frac{R}{1+\tfrac{1}{4}M(R^2-1-\Sigma_+)\sin2\psi}\ ,
  \label{eq:Rbardef} \\ 
  \bM &= \frac{M}{1+\tfrac12R^2\sin2\psi}\ .
\end{align}
We likewise replace the auxiliary variables $Z$ and $\Omega$ by the
variables $\bZ$ and $\bO$, defined by
\begin{align}
  \bZ &= \frac{Z}{1+\tfrac12M(1+\Sigma_+)\sin2\psi}\ , 
  \label{eq:Zbardef}\\  
  \bO &= \frac{\Omega}{1+\tfrac12 MR^2\sin2\psi}\ .
  \label{eq:Ombardef} 
\end{align}
These new variables are defined so as to ensure that the oscillatory
terms in their evolution equations tend to zero as \tinfty. This
follows since the oscillatory terms in
(\ref{eq:speqZ})-(\ref{eq:MeqZ}) are bounded as \tinfty.
The evolution equations, which follow from
(\ref{eq:speqZ})-(\ref{eq:psieq}), (\ref{eq:Omeq}), and
(\ref{eq:Zeq}), have the form
\begin{align}
  \bS' &= -\bR^2 -
  \bS\left(1-\bS\right) +
  MR^2B_{\bar{\Sigma}_+}\ , 
  \label{eq:spbareq} \\
  \bR' &= \left[\bS + \bS^2 +
    MB_{\bR}\right]\bR\ , \label{eq:Rbareq}\\
  \bM' &= -\left[ (1+\bS)^2 +
    MB_{\bM}\right]\bM \ , 
  \label{eq:Mbareq} \\
  \bZ' &= \left[ -1 + MB_{\bZ} \right]\bZ \ ,
  \label{eq:Zbareq} \\  
  \bO' &= 2\left[\bS^2 +
    MR^2B_{\bO}\right]\bO\ , \label{eq:Ombareq} 
\end{align}
where the $B$'s are bounded functions as \tinfty. 
Equations (\ref{eq:spbareq})-(\ref{eq:Ombareq}) depend explicitly on
$\psi$ through the B's which are all multiplied by a 
factor of $M$. This fact, in conjunction with (\ref{eq:Mscaling}),
implies that the oscillatory terms in
(\ref{eq:spbareq})-(\ref{eq:Ombareq}) tend to zero as $\tinfty$, as
desired. The structure of (\ref{eq:spbareq})-(\ref{eq:Ombareq})
now allows the limits of $\bS$ and $\bR$ as \tinfty to
be found, and subsequently the corresponding limits of $\Sigma_+$ and
$R$. The main result is contained in the the following theorem.  

\begin{theo}
  \label{theo:main}
  Any solution of the differential equations
  (\ref{eq:speqZ})-(\ref{eq:psieq}) with $\Omega>0$ that is not LRS,
  satisfies (\ref{eq:Mscaling}) and
  $$
  \lim_{\tinfty}\Sigma_+ = 0 \ , \quad
  \lim_{\tinfty} R = 0\ ,
  $$
  with
  $$
  \lim_{\tinfty} \frac{M}{R} = 0 \ .
  $$
\end{theo}
{\bf Proof:} The proof, which is based on equations
(\ref{eq:spbareq})-(\ref{eq:Ombareq}), is given in
Appendix A.\\  
\qed\\
\\
In the next Section we will consider the physical implications of 
this theorem as regards isotropization of the cosmological model.

\section{Anisotropy of the models at late times}
\label{sec:anisotropy}

To quantify the deviation of non--tilted Bianchi \VII models from
the flat FL model, we introduce a set of dimensionless {\em anisotropy 
scalars}, $\left\{ \Sigma, {\cal S}, {\cal W} \right\}$, all of
which are zero for any FL model. The anisotropy in the 
expansion of a cosmological model is described by the
{\em shear scalar} $\Sigma$, defined by 
\begin{equation}
  \Sigma^2 = \frac{\sigma_{ab}\sigma^{ab}}{6H^2}\ ,
\end{equation}
where $\sigma_{ab}$ is the rate of shear tensor of the fluid
congruence\footnote{It is this tensor that is parametrized by the
  variables $\sigma_\pm$ in (\ref{eq:gurka}).}. The anisotropy in the
spatial geometry is described by the scalar ${\cal S}$, defined by
\begin{equation}
  {\cal S}^2 = \frac{^3S_{ab}\mbox{}^3S^{ab}}{6H^2}\ ,
\end{equation}
where $^3S_{ab}$ is the trace-free spatial Ricci tensor (see
Wainwright \& Ellis (1997), page
29). Finally, the intrinsic anisotropy of the gravitational field is
described by the scalar ${\cal W}$, defined by
\begin{equation}
  {\cal W}^2 = \frac{E_{ab}E^{ab} + H_{ab}H^{ab}}{6H^2}\ , 
\end{equation}
where $E_{ab}$ and $H_{ab}$ are the electric and magnetic part of the
Weyl tensor with respect to the fluid congruence respectively (see
Wainwright \& Ellis (1997),
page 19). The scalars ${\cal S}$ and ${\cal W}$ describe the dynamical
importance of the spatial curvature and the Weyl curvature compared to
the overall expansion of the universe respectively. The use of
expansion--normalized anisotropy 
scalars dates back to the important paper on cosmological observations
by Kristian \& Sachs (1966) (see page 398), where they showed that
various geometrical quantities, including the Weyl tensor can, in
principle, be restricted by observations. For non--tilted Bianchi
models, a zero shear tensor, $\sigma_{ab}=0$, implies that the 
Weyl tensor is zero, and can thus be viewed as characterizing the FL
models. Restricting the expansion--normalized shear scalar $\Sigma$ to
be small, however, 
does not guarantee that the expansion--normalized Weyl tensor, as
described by ${\cal W}$, is small. For a model to be
close to an FL model as \tinfty, we therefore require that
\begin{equation}
  \label{eq:closetoFL}
  \Sigma \ll 1\ , \quad {\cal S} \ll 1\ , \quad {\cal W} \ll 1\ ,
\end{equation}
as \tinfty. The limits of the anisotropy scalars $\left\{ \Sigma, 
  {\cal S}, {\cal W} \right\}$ as \tinfty for non-tilted Bianchi \VII
models with a radiation fluid are given by the following theorem.  
\begin{theo}
  For any non--tilted radiation--filled Bianchi \VII cosmological
  model that is not LRS, the density parameter satisfies
  $$
  \lim_{\tinfty} \Omega = 1\ ,
  $$
  and the anisotropy scalars satisfy
  $$
  \lim_{\tinfty} \Sigma = 0 \ , \quad
  \lim_{\tinfty} {\cal W} = +\infty\ ,
  $$
  and
  $$\limsup_{\tinfty} {\cal S} = +\infty\ , \quad
  \liminf_{\tinfty} {\cal S} = 0\ .
  $$
\end{theo}
{\bf Proof:} This theorem is an immediate consequence of Theorem 1,
once $\Sigma$, ${\cal S}$, and ${\cal W}$ are expressed in terms of
the variables $\Sigma_+, R,M$, and $\psi$. In WHU it was shown that 
\begin{equation}
  \label{eq:Sigasdef}
  \Sigma^2 = \Sigma_+^2 + R^2\cos^2\psi\ ,
\end{equation}
and
\begin{equation}
  \label{eq:Wasdef}
  {\cal W} = \frac{2R}{M}\left[ 1+ \bigO{M} \right]\ ,
\end{equation}
(see WHU, equations (3.33) and (3.39)). Since $^3S_{ab}$ is
diagonal for Bianchi type \VII models, it follows that
\begin{equation}
  {\cal S}^2 = {\cal S}_+^2 + {\cal S}_-^2\ ,
\end{equation}
(see Wainwright \& Ellis (1997), page 123), where
\begin{equation}
  {\cal S}_+ = \frac{1}{2H}\left( ^3S_{22} + \mbox{}^3S_{33} \right)\
  , \quad 
  {\cal S}_- = \frac{\sqrt{3}}{2H}\left( ^3S_{22} - \mbox{}^3S_{33}
  \right)\ 
  .
\end{equation}
Specializing equation (6.36) in Wainwright \& Ellis (1997) to Bianchi
\VII models yields 
\begin{equation}
  {\cal S}_+ = 2N_-N_+\, \quad {\cal S}_- = 2N_-^2\ , 
\end{equation}
which, in conjunction with (\ref{eq:newvarintro}), implies that
\begin{equation}
  \label{eq:Sasdef}
  {\cal S}^2 = \frac{4R^2\sin^2\psi}{M^2}\left[1+\bigO{M^2}\right]\ . 
\end{equation}
Equations (\ref{eq:Sigasdef}), (\ref{eq:Wasdef}), and
(\ref{eq:Sasdef}) and Theorem 1 now give Theorem
2.\\ 
\qed\\

The limits in Theorem 2 show that radiation filled Bianchi \VII models
isotropize with respect to the shear, but not with respect to the
spatial curvature or the Weyl tensor. In view of
(\ref{eq:closetoFL}), therefore, a general non-tilted Bianchi \VII
model with a radiation-fluid source is {\em not close to any FL model
  at late times}.

\section{The asymptotic solution at late times}
\label{sec:decay}

In this Section we determine the asymptotic form of the gravitational
field variables $\Sigma_+$, $R$, $M$, and $\psi$ as \tinfty. From
these expressions, the asymptotic forms of the anisotropy scalars of
the previous Section can subsequently be found. Theorem
1, in conjunction with (\ref{eq:Mbareq}), implies that 
\begin{equation}
  \label{eq:riverworld}
  M = \bigO{{\rm e}^{-b\tau}}\ ,
\end{equation}
as \tinfty, where the constant $b$ can be written as
$b=1-\delta$. Here the constant $\delta>0$ can be
chosen arbitrarily small. It now follows from (\ref{eq:Zbareq}) that
there exists a constant $C_{Z}$ such that 
$\bZ = C_{Z}{\rm e}^{-\tau}\left[ 1+\bigO{{\rm e}^{-b\tau}}
\right]$. Hence, using (\ref{eq:Zbardef}), we find 
\begin{equation}
\label{eq:fabulous}
  Z = C_{Z}{\rm e}^{-\tau}\left[ 1+\bigO{{\rm e}^{-bt}}\right] \ ,
\end{equation}
as \tinfty. The asymptotic behaviour of
$\bS$ and $\bR$, and hence of $\Sigma_+$ and $R$, is
governed by (\ref{eq:spbareq}) and (\ref{eq:Rbareq}). We begin our
analysis by considering the differential equations in ${\mathbb R}^2$
obtained by omitting the terms in (\ref{eq:spbareq}) and
(\ref{eq:Rbareq}) that involve $M$. The resulting differential
equations are 
\begin{align}
  \hat{R}' &= \hat{\Sigma}_+(1+\hat{\Sigma}_+)\hat{R}\ ,
  \label{eq:Rredeq} \\ 
  \hat{\Sigma}_+' &= - \hat{\Sigma}_+(1-\hat{\Sigma}_+^2) - \hat{R}^2\
  ,  \label{eq:Sredeq} 
\end{align}
which we will refer to as the {\em truncated differential
  equations}. Their solutions will be denoted by  
$\hat{\Sigma}_+$ and $\hat{R}$. Since the proof in Appendix
  A that any solution of (\ref{eq:spbareq}) and (\ref{eq:Rbareq}) with  
$\bR>0$ and $\bO>0$ satisfies $\lim_{\tinfty}
\bS = 0 = \lim_{\tinfty} \bR$ also applies to the
truncated differential equtions (\ref{eq:Rredeq}) and
(\ref{eq:Sredeq}), it follows that 
\begin{equation}
  \lim_{\tinfty} \hat{\Sigma}_+ = 0 \ , \quad \lim_{\tinfty} \hat{R} =
  0 \ , 
\end{equation}
for any solution with $\hat{R}>0$ and
$\hat{\Omega}=1-\hat{\Sigma}_+^2-\hat{R}^2>0$. 
In other words, the origin is the future attractor for the truncated
equations (\ref{eq:Rredeq}) and (\ref{eq:Sredeq}). A straightforward
calculation shows that this equilibrium point is non--hyperbolic, with
eigenvalues $\lambda_{\hat{\Sigma}_+}=-1$ and
$\lambda_{\hat{R}}=0$. It will therefore be necessary to consider centre
manifold theory\footnote{For an introduction to centre manifold
theory, see, for example, Carr (1981).} when determining the
asymptotic forms of $\hat{\Sigma}_+$ and $\hat{R}$ as \tinfty. The
results are summarized in the following theorem.

\begin{theo}
  For any solution of the truncated differential equations
  (\ref{eq:Rredeq}) and (\ref{eq:Sredeq}) with $\hat{\Omega}>0$, and
  $\hat{R}>0$, there exists a constant $C_{R}$ such that 
  \begin{align}
    \hat{R} &= \frac{1}{\sqrt{2\tau}}\left[1-\frac{1}{4\tau}\left(\ln\tau +
      C_{{R}}\right) + \bigO{T^2}\right]\ , \label{eq:Rdecay} \\
\hat{\Sigma}_+&=
-\frac{1}{2\tau}\left[1-\frac{1}{2\tau}\left(\ln\tau + 
    C_{{R}}-2\right) +
  \bigO{T^2}\right]\ , \label{eq:Spdecay}
\end{align}
where $T=\tfrac{\ln \tau}{\tau}$.
\end{theo}

{\bf Proof:} The proof is given in Appendix B.\\
\qed\\
\\
The next step is to relate the solutions of (\ref{eq:spbareq}) and 
(\ref{eq:Rbareq}) to the asymptotic forms in Theorem 3. An important
consequence of Theorem 3 is that 
$\hat{\Sigma}_+$ {\em and} $\hat{R}$ {\em exhibit a power--law decay
  to zero}. This is the typical behaviour for non-hyperbolic
equilibrium points. Since  
(\ref{eq:spbareq}) and (\ref{eq:Rbareq}) differ from the
truncated differential equations (\ref{eq:Rredeq}) and
(\ref{eq:Sredeq}) by the addition of exponentially decaying terms, it
is plausible that the asymptotic decay rates of the solutions will not
be affected. We therefore conjecture that any solution $(\bR,
\bS)$ of  (\ref{eq:spbareq}) and 
(\ref{eq:Rbareq}) with $\bR>0$ and $\bO>0$ will have the
same asymptotic form as the solutions $(\hat{R},
\hat{\Sigma}_+)$ of the truncated differential equations
(\ref{eq:Rredeq}) and (\ref{eq:Sredeq}) as given in Theorem 3, {\em
  i.e.}, 
\begin{align}
  \bR &= \frac{1}{\sqrt{2\tau}}\left[1-\frac{1}{4\tau}\left(\ln\tau +
      C_{R}\right) + \bigO{T^2}\right] \ , \label{eq:Rnunadecay} \\ 
\bS &= -\frac{1}{2\tau}\left[1-\frac{1}{2\tau}\left(\ln\tau +  
    C_{R}-2\right) + \bigO{T^2}\right] \label{eq:Spnunadecay} \ .
\end{align}
The dependence of the constant $C_R$ on the initial conditions will be
different than that of the solution
(\ref{eq:Rdecay})-(\ref{eq:Spdecay}). We 
have not been able to give a proof of the above conjecture in general,
but in Appendix C we give a proof 
of a related but simpler result. Assuming
(\ref{eq:Rnunadecay}) and (\ref{eq:Spnunadecay}), it follows from
(\ref{eq:spbardef}), (\ref{eq:Rbardef}), and (\ref{eq:riverworld})
that 
\begin{align}
  R &= \frac{1}{\sqrt{2\tau}}\left[1-\frac{1}{4\tau}\left(\ln\tau +
      C_{R}\right) + \bigO{T^2}\right]
\ , \label{eq:Rnuna2decay} \\
\Sigma_+&=
-\frac{1}{2\tau}\left[1-\frac{1}{2\tau}\left(\ln\tau + 
    C_{R}-2\right) +
  \bigO{T^2}\right] \label{eq:Spnuna2decay} \ .
\end{align}
The asymptotic form of $M$ can be obtained from (\ref{eq:Rnuna2decay})
and (\ref{eq:Spnuna2decay}) using (\ref{eq:Zdef}) and
(\ref{eq:fabulous}). The result is
\begin{equation}
  \label{eq:Mnuna2decay}
  M = C_M\tau{\rm e}^{-\tau}\left[1 + \frac{1}{4\tau}\left(2\ln\tau +
  2C_R-1\right)  + \bigO{T^2}\right]\ ,
\end{equation}
where $C_M=2C_Z$. Subsequently, by using (\ref{eq:psieq}), we find
\begin{equation}
  \label{eq:phiapproxxx}
  \psi' = \frac{2{\rm
  e}^{\tau}}{C_M\tau}\left[1-\tfrac{1}{4\tau}(2\ln\tau +2C_R-1) +
  \bigO{T^2}\right]\ .
\end{equation}
This equation determines $\psi$ up to an additive constant $C_\psi$,
although $\psi$ cannot be expressed in elementary terms.

The Hubble variable $H$ can be determined 
algebraically through the relation $\Omega=\tfrac{\mu}{3H^2}$, since,
for a radiation fluid, the quantity $\mu l^4$ is a constant. Here $l$
is the length scale 
function, which is related to $\tau$ according to 
\begin{equation}
  \label{eq:lldef}
  l=C_l{\rm e}^\tau\ ,
\end{equation}
where $C_l$ is a constant (see WHU, equation (1.3)). It follows, using
(\ref{eq:omdef}), 
(\ref{eq:Rnuna2decay}), (\ref{eq:Spnuna2decay}), and (\ref{eq:lldef})
that $H$ is given by
\begin{equation}
  \label{eq:landH}
  H = C_H{\rm e}^{-2\tau}\left[ 1+
    \frac{1}{4\tau} + \bigO{\tfrac{\ln\tau}{\tau^2}} \right]\ ,
\end{equation}
where $C_H$ is a constant. Finally, by using
(\ref{eq:landH}), in conjunction with
(\ref{eq:introtau}), the relationship between the cosmological clock
time 
$t$ and the Hubble time $\tau$ can be found. By choosing appropriate 
initial conditions for $t$, we obtain 
\begin{equation}
  \label{eq:texpression}
  t = \frac{1}{2C_H}{\rm e}^{2\tau}\left[ 1 - \frac{1}{4\tau} +
  \bigO{\tfrac{\ln\tau}{\tau^2}}\right]\ ,
\end{equation}
as \tinfty. The freedom to shift the
time-variable according to  $\tau \rightarrow\tau+C$,
where $C$ is a constant, can be used to fix one of the constants, say
$C_l=1$. There thus remain four essential constants of integration,
namely  $C_R, C_\psi, C_M$, and $C_H$. This confirms that
(\ref{eq:Rnuna2decay})-(\ref{eq:texpression}) describe the asymptotic
form of the general Bianchi \VII solution with a radiation fluid
source.

We conclude this Section by giving the asymptotic form of the various
scalars of physical interest in terms of the clock time $t$. The
asymptotic form of the anisotropy scalars $\Sigma^2$, ${\cal W}^2$ and
${\cal S}^2$ follow immediately from (\ref{eq:Sigasdef}),
(\ref{eq:Wasdef}), (\ref{eq:Rnuna2decay})-(\ref{eq:phiapproxxx}) and
(\ref{eq:texpression}). The asymptotic form of the Hubble variable $H$
and the radiation density $\mu$ are obtained from 
(\ref{eq:state}), (\ref{eq:NewOmdef}), (\ref{eq:Rnuna2decay}),
(\ref{eq:Spnuna2decay}), (\ref{eq:landH}) and
(\ref{eq:texpression}). For convenience we introduce new constants $A$
and $\alpha$ according to
\begin{equation}
  A=2C_H\ , \quad \alpha=\frac{4}{C_M}\ .
\end{equation}
Note that $A$ has dimension of $(time)^{-1}$, while $\alpha$ is
dimensionless. The results are
\begin{align}
  H &\approx \frac{1}{2t}\left(1+\frac{1}{2\ln(At)}\right)\ ,
  \label{eq:egghead} \\ 
  \mu &\approx \frac{3}{4t^2}\left(1-\frac{1}{\ln(At)}\right)\ ,
  \label{eq:muhead}  \\
  \Sigma^2 &\approx \frac{\cos^2\psi}{\ln(At)} +
  \frac{1}{\left[\ln(At)\right]^2} \ , \label{eq:hauntedhouse}\\
  {\cal W}^2 &\approx \frac{\alpha(At)}{\left[\ln(At)\right]^3}\ ,
  \label{eq:nevecampbell}\\ 
  {\cal S}^2 &\approx {\cal W}^2\sin^2\psi \label{eq:nerdo}\ ,
\end{align}
where
\begin{equation}
  \label{eq:CourtneyCox}
  \frac{d\psi}{dt} \approx \frac{\alpha A\sqrt{At}}{2\ln(At)}\ .
\end{equation}
It is understood that these asymptotic forms are valid for values of
$A$ and $t$ such that $\ln(At)\gg 1$. As mentioned in Section
\ref{sec:intro}, these asymptotic forms could describe the dynamics of
the universe in 
the final stages of the radiation-dominated epoch, which ends at a
time $t=t_{\rm eq}$, when the matter density first equals the
radiation density. In order to describe this situation, the constant
$A$ must satisfy
\begin{equation}
  \ln(At_{\rm eq}) \gg 1\ .
\end{equation}
It follows from (\ref{eq:hauntedhouse}), in conjunction with
(\ref{eq:CourtneyCox}), that the constant $\alpha$ 
determines how rapidly $\Sigma$ oscillates. It also determines the
magnitude of the Weyl scalar ${\cal W}$ at $t=t_{\rm eq}$ (see
(\ref{eq:nevecampbell})). We shall discuss the
significance of the above results in Section 7. 

\section{Relation with the metric approach}
\label{sec:metric}

The late time behaviour of Bianchi \VII models with a radiation--fluid
source has been considered by Doroshkevich \etal (1973) using metric 
variables. In our approach, the asymptotic behaviour of the metric
components can be 
found directly from the behaviour of the orthonormal frame variables
in (\ref{eq:Rnuna2decay})-(\ref{eq:landH}), and we now do this
for the purpose of comparison.  

For a Bianchi type \VII model, a set of group--invariant
and time--independent 1--forms, $\boldsymbol{\omega}^\alpha$,
$\alpha=1,2,3$ satisfying
\begin{equation}
  d\boldsymbol{\omega}^1 = 0 \ , \quad d\boldsymbol{\omega}^2 =
  \boldsymbol{\omega}^3 \wedge \boldsymbol{\omega}^1 \ , \quad
  d\boldsymbol{\omega}^3 = 
  \boldsymbol{\omega}^1 \wedge \boldsymbol{\omega}^2\ ,
\end{equation}
can be introduced for which the line element can be
written
\begin{equation}
  \label{eq:dsDoro}
  ds^2 = -dt^2 +
  g_{\alpha\beta}\boldsymbol{\omega}^\alpha\boldsymbol{\omega}^\beta\
  .
\end{equation}
Here $t$ is (as previously) the cosmological clock time. For a 
non-tilted Bianchi \VII model, $g_{\alpha\beta}$ can be
diagonalized. Using the variables of Misner (1969) to parametrize 
$g_{\alpha\beta}$, the line element (\ref{eq:dsDoro}) can be
written as 
\begin{equation}
  \label{eq:DLNmetric}
  ds^2 = -dt^2 + l^2\left[ {\rm
  e}^{-4\beta^+}\left(\boldsymbol{\omega}^1\right)^2  + {\rm
  e}^{2(\beta^+ +
  \sqrt{3}\beta^-)}\left(\boldsymbol{\omega}^2\right)^2 + {\rm
  e}^{2(\beta^+ -
  \sqrt{3}\beta^-)}\left(\boldsymbol{\omega}^3\right)^2 \right] \ .
\end{equation}
The relationships between the metric variables $\beta^\pm$ and the
orthonormal frame variables for Bianchi class A models are given in
chapter 9 of Wainwright \& Ellis (1997). Specializing these
relationships to Bianchi \VII models yields
\begin{equation}
  N_+ + \sqrt{3}N_- = \frac{1}{Hl}{\rm e}^{2(\beta^+ +
  \sqrt{3}\beta^-)}\ , \quad N_+ - \sqrt{3}N_- = \frac{1}{Hl}{\rm
  e}^{2(\beta^+ - \sqrt{3}\beta^-)} \ .
\end{equation}
Isolating $\beta^+$ and $\beta^-$ leads to the following expressions 
\begin{equation}
  \label{eq:betarels}
  \tanh (2\sqrt{3}\beta^-) = \sqrt{3}RM\sin\psi\ , \quad 
  {\rm e}^{4\beta^+} = \frac{l^2H^2}{M^2}(1-3R^2M^2\sin^2\psi)\ ,
\end{equation}
(see Appendix C in WHU). Using (\ref{eq:Rnuna2decay}),
(\ref{eq:Spnuna2decay}), (\ref{eq:Mnuna2decay}), and
(\ref{eq:phiapproxxx}), it follows that  
\begin{align}
  \beta^+ &\approx -\tfrac12\ln\tau - \frac{1}{4\tau}\ln \tau -
  \tfrac{1}{4\tau}(C_R-1)\ , \label{eq:asympplus}\\
  \beta^- &\approx \tfrac{C_HC_l}{2\sqrt{2}}{\rm
  e}^{-\tau}\sqrt{\tau}\left[ 1+ 
  \frac{1}{4\tau}\ln\tau + \frac{1}{4\tau}(C_R-1)\right]\sin\psi
  \label{eq:asympminus} \ .
\end{align}
To facilitate a comparison with the asymptotic expressions of
Doroshkevich \etal (1973), we note that their variables $\gamma, \mu$
and $(\lambda_1\lambda_2)^{1/2}$ are related to the variables used in
this paper by
\begin{equation}
  \gamma=l^6\ , \quad \mu=4\sqrt{3}\beta^-\ , \quad 
  (\lambda_1\lambda_2)^{1/2}=l^2{\rm e}^{2\beta^+}\ ,
\end{equation}
and that their time variable $T$ (the so--called BKL time) is related
to $\tau$ by  
\begin{equation}
  \label{eq:times}
  \frac{d\tau}{dT} = Hl^3 \quad \Rightarrow \quad T \approx
  -\left(C_HC_l\right)^{-1} {\rm e}^{-\tau}\left( 1-\frac{1}{4\tau}
  \right) \ .
\end{equation}
We note that $T$ approaches zero from below as \tinfty. 
Doroshkevich \etal (1973) give the following asymptotic forms:
\begin{equation}
  \label{eq:dorobetas}
  l^2{\rm e}^{2\beta^+} \approx D_1\frac{\theta(T)}{T^2} \ , \quad
  \beta^- \approx D_2\theta(T)^{-1/2}T\sin\psi(T)\ , \quad
  \frac{d\psi}{dT} \approx \frac{D_3\theta(T)}{T^2}\ ,
\end{equation}
see Appendix I, equation (I.14). Note that we have relabeled their
constants. The function $\theta(T)$ in Doroshkevich \etal (1973)
is given by 
\begin{equation}
  \label{eq:thetadef}
  \theta(T) \approx \frac{1}{\nu}\left[1-\frac{D_4}{\nu} -
  \frac{\ln\nu}{2\nu}\right] \ , 
\end{equation}
where $\nu = \ln(D_5/T)$ (see Appendix I, equation (I.15) ). Using
(\ref{eq:times}) in (\ref{eq:thetadef}) yields
\begin{equation}
  \label{eq:thetaapprox}
  \theta \approx \frac{1}{\tau}\left( 1-\frac{D_*}{\tau} -
  \frac{\ln\tau}{2\tau}\right) \ , 
\end{equation}
where the constant $D_*$ is a combination of the constants $C_H,
C_l, D_4$ and $D_5$. Using (\ref{eq:times}) and (\ref{eq:thetaapprox})
in (\ref{eq:dorobetas}) yields (\ref{eq:asympplus}) and
(\ref{eq:asympminus}) if we identify $C_R=2D_*$,
$D_1=C_H^{-2}C_l^{-4}$, and $D_2=-\sqrt{6}/D_1$. From
(\ref{eq:times}) it also follows that the evolution equation for
$\psi$ in (\ref{eq:dorobetas}) is (asymptotically) equivalent to
(\ref{eq:phiapproxxx}) if we identify $D_3=2/D_1$. 

\section{Discussion}
\label{sec:disc}

In this paper we have given a complete analysis of the behaviour at
late times of non-tilted Bianchi \VII cosmologies filled with
radiation. In particular, we have established rigorously the late time
behaviour of the dimensionless anisotropy scalars $\Sigma, {\cal W}$,
and ${\cal S}$, that determine the extent to which the models deviate
from an FL model (Theorem 2). In addition, we have derived the
asymptotic form as \tinfty of the general solution, although in this
regard we have had to rely on a heuristic argument at one stage of the
analysis.

Our results show that the Bianchi \VII radiation models differ from
the dust models in two significant ways:
\begin{enumerate}
  \item for dust models, the dimensionless Weyl scalar ${\cal W}$ has
  a finite non-zero limit (see Theorem 2.4 in WHU), while for
  radiation-filled 
  models, ${\cal W}$ increases without bound,
  \item for dust models, the dimensionless shear scalar $\Sigma$ tends
  to zero at an exponential rate in terms of the Hubble time $\tau$,
  while for radiation models $\Sigma$ decays at a power law rate in
  terms of $\tau$. 
\end{enumerate}

The results also shed new light on the dynamics of the radiation
models at 
late times. These models isotropize as regards the shear ({\em i.e.}
$\Sigma\rightarrow 0$), but not as regards the Weyl curvature and the
anisotropic spatial curvature ({\em i.e.} ${\cal W} \nrightarrow 0,
{\cal S} \nrightarrow 0$). Equations
(\ref{eq:egghead})-(\ref{eq:nerdo}) give 
the asymptotic dependence of the various scalars of physical interest
on the clock time $t$. We note that the leading order terms of
$H$ and $\mu$ in (\ref{eq:egghead}) and (\ref{eq:muhead}) give the
flat FL radiation solution, and that, in addition, the shear
scalar $\Sigma$ in 
(\ref{eq:hauntedhouse}) tends to zero. The behaviour of ${\cal W}$ 
and ${\cal S}$ in (\ref{eq:nevecampbell})
and (\ref{eq:nerdo}), however, implies that the solution is {\em not}
close to the flat FL model at late times. Moreover, since ${\cal W}$
diverges at late times, the 
models exhibit the phenomenon of {\em Weyl curvature dominance}, {\em
  i.e.}, the Weyl curvature plays a dominant role in determining the
dynamics. These considerations show that solving the
linearized Einstein field equations can lead to incorrect conclusions
regarding the dynamics of the full equations. It was pointed out in
WHU (see pages 2588-89) that some aspects of the late time evolution
of dust models of Bianchi \VII are correctly described by linearizing
the Bianchi \VII evolution equations about the flat FL model. On the
other hand, {\em 
  for radiation models, the linearized solutions totally fail to
  describe the asymptotic behaviour of the shear, the Weyl curvature,
  and the anisotropic spatial curvature at late times\footnote{Compare
    equation (4.3) in WHU, setting $\beta=0$ ({\rm i.e.}
    $\gamma=\tfrac{4}{3}$), with our equations (46)-(48).}, implying
  that this behaviour is a fully non-linear phenomenon}.   

The rate at which the shear scalar $\Sigma$ tends to zero requires
further comment. Equation (\ref{eq:hauntedhouse}) shows that the
asymptotic form of 
$\Sigma^2$ is the sum of two terms. One of
them is oscillatory and decays as $(\ln At)^{-1}$, while the other is
non-oscillatory and decays as $(\ln At)^{-2}$. It appears that the
dominant 
oscillatory term has been inadvertently 
omitted in the literature giving the incorrect result that $\Sigma
\approx \tfrac{1}{\ln(At)}$ as $t\rightarrow\infty$. For example,
Doroshkevich \etal (1973) state that ``the anisotropy of the
deformation tensor, ..., decreases very slowly, proportionally to $(\ln
t)^{-1}$'' (see page 741), but do not specify what normalization was
used to calculate the ``anisotropy of the deformation tensor''. In a
subsequent article, however, Novikov (1974), states the conclusion
more explicitly, namely that the components of the shear tensor, when
normalized with the Hubble scalar $H$, decay as
$\left[\ln(t/t_F)\right]^{-1}$, (see equation (2) on page 275, and the
definition\footnote{We note that our shear variables $\Sigma_\pm$ are
  linear combinations of the $\Delta H_i/H$ in Novikov (1974).} of
$\Delta H_i/H$ on 
page 274). This incorrect result was subsequently quoted by Barrow
(1976) (see the equation for $\sigma/\theta$ in the case
$p=\tfrac{1}{3}\mu$ , on page 364). 

As mentioned in Section 1, the late time behaviour if Bianchi \VII
radiation models could describe the dynamics of the universe in the
final stages of the radiation dominated epoch. Our results show that
on transition from the radiation-dominated epoch to the
matter-dominated epoch at $t=t_{\rm eq}$, the shear scalar would be
small but the Weyl curvature scalar ${\cal W}$ would be large. These
values of $\Sigma$ and ${\cal W}$ would then act as initial values, so
to 
speak, for the matter-dominated epoch. Since $\Sigma\rightarrow 0$ and
${\cal W}\rightarrow {\cal W}_\infty$, where ${\cal W}_\infty$ is a
positive constant, as $t\rightarrow\infty$ for a dust model, the shear
scalar would remain small, but the 
dynamical significance of ${\cal W}$ would persist. It would be of
interest to investigate density perturbations in such a model, since
they could well behave differently than in a perturbed FL model.

\section*{Acknowledgements}
The authors wish to thank David Siegel for supplying the proof of
Proposition 1 in Appendix C. The research was supported in part by a
grant from the Natural Sciences \& Engineering Research Council of
Canada (JW), G{\aa}l\"ostiftelsen (USN), Svenska Institutet (USN),
Stiftelsen Blanceflor (USN) and the University of Waterloo (USN, MH).

\section*{Appendix A: Proof of Theorem \ref{theo:main}}
In this Appendix we give the proof of Theorem \ref{theo:main}. The
proof is based on the evolution equation for $\bar{\Omega}$ and the
fact that $\bar{\Omega}$ is bounded. We begin by integrating equation 
(\ref{eq:Ombareq}) to obtain
\begin{equation}
  \label{eq:Ombarint}
  \tfrac12\ln\frac{\bar{\Omega}}{\bar{\Omega}_0} = \int_{\tau_0}^\tau
  \bar{\Sigma}_+^2 ds + \int_{\tau_0}^\tau
  M\bar{R}^2B_{\bar{\Omega}}ds\ , 
\end{equation}
where $\bar{\Omega}_0 = \bar{\Omega}(\tau_0)$ and the initial time
$\tau_0$ is chosen such that $M\leq1$ for all 
$\tau\geq\tau_0$. Equation (\ref{eq:Zbareq}) implies that
\begin{equation}
  \label{eq:Zscaling}
  \bar{Z} = \bigO{{\rm e}^{(-1+\delta)\tau}}\ ,
\end{equation}
where $\delta$ can be chosen arbitrarily small\footnote{Since $M$
  tends to zero and $B_{\bar{Z}}$ is bounded, one can choose $\tau_0$
  such that $\left|MB_1\right|<\delta$ for $\tau>\tau_0$.}. We now
require the fact that $\bar{\Omega}$ is bounded, which follows from
(\ref{eq:NewOmdef}), (\ref{eq:Mscaling}), and
(\ref{eq:Ombardef}). Recalling the definitions (\ref{eq:Zdef}) and
(\ref{eq:Zbardef}), equation (\ref{eq:Zscaling}) implies
\begin{equation}
  \label{eq:mr2scaling}
  \bar{M}\bar{R}^2 = \bigO{{\rm e}^{(-1+\delta)\tau}}\ ,
\end{equation}
and hence
\begin{equation}
  \label{eq:B4int}
  \int_{\tau_0}^\tau M\bar{R}^2B_{\bar{\Omega}}ds = C + \bigO{{\rm
  e}^{(-1 + \delta)\tau}}\ ,
\end{equation}
where $C$ is a constant depending on the initial conditions at
$\tau=\tau_0$. Using (\ref{eq:B4int}) and the fact that $\Omega$ is
positive and bounded, we conclude from (\ref{eq:Ombarint}) that
$I(\tau) = \int_{\tau_0}^\tau 
\bar{\Sigma}_+^2 ds$ is bounded above for all
$\tau\geq\tau_0$. Since the integrand is non-negative, the function
$I(\tau)$ is monotonically decreasing, which implies that 
$\lim_{\tinfty} I(\tau)$ exists and is finite. It
follows from (\ref{eq:Ombarint}), (\ref{eq:B4int}) and
(\ref{eq:Ombardef}) that 
$\lim_{\tinfty} \bar{\Omega}$, and hence
$\lim_{\tinfty} \Omega$ exists, \ie there is a constant
$L$, satisfying $0\leq L\leq1$, such that
\begin{equation}
  \label{eq:Ombarlimit}
  \lim_{\tau\rightarrow\infty}{\Omega} = L\ .
\end{equation}
In order to proceed, we need the following standard result
\begin{lemma}
  \label{lemma:splimit}
  Let $f(\tau)$ be a non-negative real-valued and Lipschitz continuous
  function on the interval $\tau_0\leq\tau<+\infty$. If
  $$
  \int_{\tau_0}^\infty f(\tau)d\tau
  $$
  is finite then $\lim_{\tinfty}f(\tau)=0$.
\end{lemma}
Since $\bS$ and $\bR$ are bounded, it follows from (\ref{eq:spbareq})
that $\left(\bS^2\right)' = 2\bS\bS'$ is uniformly bounded for
$\tau\geq\tau_0$. This implies that $\bS$ is Lipschitz continuous on
the interval $\tau\geq\tau_0$ (see ???). We may now apply Lemma
\ref{lemma:splimit}, with $f=\bS^2$, to conclude that
\begin{equation}
  \label{eq:groucho}
  \lim_{\tinfty}\bS=0\ ,
\end{equation}
and hence, from (\ref{eq:Mscaling}) and (\ref{eq:spbardef})
\begin{equation}
  \label{eq:spbarlimit}
  \lim_{\tinfty} \Sigma_+ = 0\ .
\end{equation}
From (\ref{eq:omdef}), (\ref{eq:Rbardef}), (\ref{eq:Ombarlimit}), and 
(\ref{eq:spbarlimit}), it follows that 
\begin{equation}
  \label{eq:barRlimit}
  \lim_{\tinfty} R^2 = \lim_{\tinfty}
  \bR^2 = 1-L\ .
\end{equation}
Equation (\ref{eq:spbareq}), in conjunction with
(\ref{eq:spbarlimit}), (\ref{eq:barRlimit}), and (\ref{eq:Mscaling}),
now implies that 
$\lim_{\tinfty} \bS'$ exists, and that
\begin{equation}
  \lim_{\tinfty}\bar{\Sigma}_+' = L-1\ .
\end{equation}
This result contradicts (\ref{eq:groucho}) unless 
$L=1$. Using (\ref{eq:barRlimit}) we obtain
\begin{equation}
  \lim_{\tinfty} R = 0 \ . 
\end{equation}
To find the limit of the quotient $\tfrac{M}{R}$, we note that
(\ref{eq:Rbareq}) and (\ref{eq:Mbareq}) can be combined to give
\begin{equation}
  \left( \frac{\bar{M}}{\bar{R}} \right)' = \left(-1-3\bar{\Sigma}_+ -
  2\bar{\Sigma}_+^2 + MB_*\right)\left(
  \frac{\bar{M}}{\bar{R}} \right)\ ,
\end{equation}
where $B_*$ is a bounded function as
\tinfty. It now follows from (\ref{eq:groucho}) and (\ref{eq:Mscaling})
that
\begin{equation}
  \lim_{\tinfty} \frac{\bar{M}}{\bar{R}} = 0\ ,
\end{equation}
and hence that
\begin{equation}
  \lim_{\tinfty} \frac{M}{R} = 0\ .
\end{equation}
\qed

\section*{Appendix B: Proof of theorem 3}
In this Appendix we give the proof of Theorem 3, using centre manifold
theory. In developing this theory, Carr (1981) (see his
equation (2.3.1)), considers a system of differential equations of the
form
\begin{align}
  x' &= Ax + f( x, y )\ ,
  \label{eq:xeq} \\ 
  y' &= By + g( x, y )\ ,
  \label{eq:yeq} 
\end{align}
where $x\in{\mathbb R}^n$, $y\in{\mathbb R}^m$, $A$ is a constant
$n\times n$ matrix, $B$ is a constant $m\times m$ matrix, and
$f:{\mathbb R}^n \times {\mathbb R}^m \rightarrow {\mathbb R}^n$ and 
$g:{\mathbb R}^n \times {\mathbb R}^m \rightarrow {\mathbb R}^m$ are $C^2$
functions that are zero and have zero derivatives at the equilibrium
point $(x,y)=(0,0)$. It is assumed that the eigenvalues of
$A$ have zero real parts and that the eigenvalues of $B$
have negative real parts. The assumption on $A$ means that the
equilibrium point $(x, y) = (0,0)$ is
non-hyperbolic, and hence that the linearization of (\ref{eq:xeq}) and
(\ref{eq:yeq}) at this equilibrium point does not describe the behaviour
of solutions near the equilibrium point.

By theorem 1 in Carr (1981), the differential equations
(\ref{eq:xeq})-(\ref{eq:yeq}) has a local center manifold 
$y=h(x)$, where $h \in C^2$, and the flow on this 
center manifold is governed by the reduced differential equation
\begin{equation}
  \label{eq:ueq}
  u' = f(u, h(u))\ ,
\end{equation}
(see equation (2.4.1) in Carr (1981)). By theorem 2 in Carr (1981) if 
the equilibrium point $u=0$ of (\ref{eq:ueq}) is
asymptotically stable then the equilibrium point $(x, y) =
(0,0)$ of (\ref{eq:xeq})-(\ref{eq:yeq}) is also asymptotically
stable. In addition, if $\left( x(t), y(t) \right)$ is a
solution of (\ref{eq:xeq})-(\ref{eq:yeq}) with $\left( x(0),
  y(0) \right)$ sufficiently close to the origin, then there
exists a solution $u(t)$ of (\ref{eq:ueq}) such that\footnote{It
  is possible that the solution $u(t)$ is the zero solution, in
  which case $x(t)$ and $y(t)$ decay exponentially to
  zero. This special case will occur only if the initial point
  $\left( x(0),y(0) \right)$  lies on the stable manifold
  of the equilibrium point $(x, y) = (0,0)$ of the
  differential equations (\ref{eq:xeq})-(\ref{eq:yeq}).}
\begin{align}
  x(t) &= u(t) + \bigO{{\rm e}^{-\gamma t}}\ , \\
  y(t) &= h( u(t) ) + \bigO{{\rm e}^{-\gamma t}}\ ,
\end{align}
where $\gamma>0$ is a constant depending only on the matrix $B$
(see Carr (1981), equation (2.4.5)). The implication of this result is
that if the equilibrium point $u=0$ of (\ref{eq:ueq}) is
asymptotically stable, one can determine the asymptotic behaviour of
the solutions of (\ref{eq:xeq})-(\ref{eq:yeq}) by finding the
asymptotic behaviour of solutions of (\ref{eq:ueq}). In order to
perform this analysis one has to approximate the center manifold 
$y=h(x)$, as follows. For a function $\phi:{\mathbb R}^n
\rightarrow {\mathbb R}^m$ that is $C^1$ near the origin, define an
operator $M$ by
\begin{equation}
  \label{eq:Moperator}
  \left(M\phi\right)(x) = \phi'(x)\left[Ax +
    f(x, \phi(x) )\right] - \left[ B\phi({
      x}) + g(x, \phi(x))\right]\ .
\end{equation}
By theorem 3 in Carr (1981), if $\phi(0)=0, \; \phi'(0)=0$ and
$\left(M\phi\right)(x)=\bigO{\left|x\right|^q}$ as $x\rightarrow 
 0$, where $q>1$, then 
\begin{equation}
  \label{eq:cfapprox}
  \left| h(x) - \phi(x)\right| = \bigO{\left|
  x\right|^q}\ ,
\end{equation}
as $x\rightarrow 0$. In other words, the function $\phi(x)$
approximates the center 
manifold $y={h}(x)$ with an accuracy determined by
$q$.

The differential equation (\ref{eq:Rredeq})-(\ref{eq:Sredeq}) is of
the form (\ref{eq:xeq})-(\ref{eq:yeq}) with $n=1$, $m=1$,
\begin{equation}
\left( x, y \right) = (\hat{R},
\hat{\Sigma}_+) \label{eq:xyinRS}\ ,
\end{equation}
\begin{equation}
  A=0\ , \quad B=1 \label{eq:ABnew}\ ,
\end{equation}
\begin{equation}
  f({x,y}) = xy(1+y)\ , \quad g(x,y) = -x^2+y^3 \label{eq:fgnew}\
  . 
\end{equation}
In addition, we have shown that the equilibrium point $(\hat{R},
  \hat{\Sigma}_+)$ is asymptotically stable, so we can apply the
above theory. Using (\ref{eq:Moperator}) and (\ref{eq:cfapprox}), in
conjunction with (\ref{eq:xyinRS})-(\ref{eq:fgnew}), we find that
the center manifold is approximated by
\begin{equation}
  \label{eq:expcf}
  \hat{\Sigma}_+ = -\hat{R}^2 - 2\hat{R}^4 + \bigO{\hat{R}^6}\ .
\end{equation}
Substituting this in (\ref{eq:Rredeq}) gives the reduced differential
equation (\ref{eq:ueq}) on the center manifold in the form
\begin{equation}
  \label{eq:Rmelin}
  \hat{R}' = -\hat{R}^3 - \hat{R}^5 + \bigO{\hat{R}^7}\ .
\end{equation}
This differential equation is precisely equation (3.1.9) with $a=-2$,
considered by Carr (1981), who shows that the asymptotic form of the
solution is
\begin{equation}
  \label{eq:carrsol}
  \hat{R} = \frac{1}{\sqrt{2\tau}}\left[1 -
  \frac{1}{4\tau}\left(\ln\tau + C_{R}\right) +
  o\left(\tau^{-1}\right)\right]\ ,
\end{equation}
as $\tinfty$, where $C_{R}$ is a constant that depends on the
initial conditions. The error bound $o(\tau^{-1})$ can be improved to
$\bigO{\left(\tfrac{\ln\tau}{\tau}\right)^2}$, as in Theorem 3, by
increasing the accuracy of the center manifold approximation
(\ref{eq:expcf}) to
\begin{equation}
  \hat{\Sigma}_+ = -\hat{R}^2 - 2\hat{R}^4 - 11\hat{R}^6 +
  \bigO{\hat{R}^8}\ . 
\end{equation}
Finally, substituting (\ref{eq:carrsol}) in
(\ref{eq:expcf}) gives the asymptotic form of $\hat{\Sigma}_+$, as
stated in Theorem 3.\\
\qed\\

\section*{Appendix C}
In this Appendix we prove that any solution to the perturbed
differential equation 
\begin{equation}
  \label{eq:fullDE}
  y' = -y^3 + \epsilon{\rm e}^{-\gamma\tau}\ ,
\end{equation}
where $\epsilon>0$ and $\gamma>0$ are constants, has the same
asymptotic form as \tinfty as the solutions to the unperturbed
differential equation  
\begin{equation}
  \label{eq:simpeDE}
  Y' = -Y^3\ .
\end{equation}
Equation (\ref{eq:simpeDE}) is a simplified form of the differential
equation (\ref{eq:Rmelin}) that describes the flow on the center
manifold in Appendix B, and the term $\epsilon{\rm e}^{-\gamma\tau}$
mimics the perturbation introduced by the exponentially decaying
function $M$.
\begin{prop}
  For any solution $y(\tau)$ of (\ref{eq:fullDE}) there exists a solution
  $Y(\tau)$ of (\ref{eq:simpeDE}) such that
  $$
  \left| y(\tau) - Y(\tau)\right| = \bigO{{\rm e}^{-\gamma\tau}}\ ,
  $$
  as \tinfty.
\end{prop}
{\bf Proof:}
Let
$y(\tau)$ be the solution of (\ref{eq:fullDE}) that satisfies the
initial condition 
$y(0)=y_0$ where $y_0>0$. The differential equation (\ref{eq:fullDE})
then implies that $y(t)$ remains positive and is bounded above. This,
in turn, implies that the solution exists for all $\tau\geq0$. 

We first need to derive an upper bound for $\tfrac{1}{y(\tau)}$. The
differential equation (\ref{eq:fullDE}) implies that
\begin{equation}
  -\frac{y'}{y^3} \leq 1\ ,
\end{equation}
which, when integrated from 0 to $\tau$, gives
\begin{equation}
  \label{eq:0tauint}
  \frac{1}{y(\tau)^2} \leq 2\tau + \frac{1}{y_0^2} \ .
\end{equation}
We now define
\begin{equation}
  \label{eq:wdefinition}
  w(\tau) = \frac{1}{2y(\tau)^2} - \tau\ .
\end{equation}
The differential equation (\ref{eq:fullDE}) implies that
\begin{equation}
  \label{eq:wdiffeq}
  w' = \frac{\epsilon{\rm e}^{-\gamma\tau}}{y(\tau)^3}\ .
\end{equation}
Integrating (\ref{eq:wdiffeq}) from 0 to $\tau$ gives
\begin{equation}
  \label{eq:wtauw0}
  w(\tau) - w(0) = -\epsilon\int_0^\tau\frac{{\rm e}^{-\gamma
  s}}{y(s)^3}ds\ .
\end{equation}
Since $\tfrac{1}{y(s)^3}$ is bounded (see (\ref{eq:0tauint})), it
follows that $\lim_{\tinfty} w(\tau)$ exists and equals $c$, where
\begin{equation}
  \label{eq:climit}
  c=w(0) - \epsilon\int_0^\infty\frac{{\rm e}^{-\gamma
  s}}{y(s)^3}ds\ .
\end{equation}
Subtracting (\ref{eq:climit}) from (\ref{eq:wtauw0}) and using
(\ref{eq:0tauint}) yields
\begin{equation}
  w(\tau) - c = \epsilon\int_\tau^\infty\frac{{\rm e}^{-\gamma
      s}}{y(s)^3}ds \leq \epsilon\int_\tau^\infty {\rm
      e}^{-\gamma s}\left( 2s + \frac{1}{y_0^2} \right)^{3/2} ds\ .
  \end{equation}
Thus,
\begin{equation}
  w(\tau) = c+ \bigO{\tau^{3/2}{\rm e}^{-\gamma\tau}}\ ,
\end{equation}
as \tinfty. It now follows from (\ref{eq:wdefinition}) that
\begin{equation}
  y(\tau) = \frac{1}{\sqrt{2}\sqrt{\tau+c+\bigO{\tau^{3/2}{\rm
  e}^{-\gamma\tau}}}}\ ,
\end{equation}
which can be rearranged to give
\begin{equation}
  y(\tau) = \frac{1}{\sqrt{2}\sqrt{\tau+c}} + \bigO{{\rm
  e}^{-\gamma\tau}} \ ,
\end{equation}
as \tinfty. Since the first term is a solution of (\ref{eq:simpeDE}),
the proof is complete.\\
\qed\\

\section*{References}

\noindent
  \hangindent=3em
Barrow J 1976 Light elements and the isotropy of the universe {\em
  Mon. Not. R. astr. Soc.} {\bf 175} 359-370

\noindent
  \hangindent=3em
  Carr J 1981 {\it Application of center manifold theory} (Springer
  Verlag: New York)

  \noindent
  \hangindent=3em
  Doroshkevich A G, Lukash V N and Novikov I D 1973 The isotropization
  of homogeneous cosmological models \jetp {\bf 37} 739-746

  \noindent
  \hangindent=3em
  Ellis G F R and MacCallum M A H 1969 A class of homogeneous
  cosmological models \cmp {\bf 12} 108-141

  \noindent
  \hangindent=3em
  Kristian J and Sachs R K 1966 Observations in cosmology  \apj {\bf
    143} 379-399

  \noindent
  \hangindent=3em
  Lukash V N 1974 Some pecularities in the evolution of homogeneous
  anisotropic cosmological models {\em Sov. Astron.} {\bf 18}
  164-169   

  \noindent
  \hangindent=3em
  Lukash V N 1975 Gravitational waves that conserve the homogeneity of
  space \jetp {\bf 40} 792-799

  \noindent
  \hangindent=3em
  Lukash V N 1976 Physical Interpretation of homogeneous cosmological
  models {\em Nuovo Cimento} B {\bf 35} 268-292

  \noindent
  \hangindent=3em
  Misner C W 1969 Quantum Cosmology I \PR {\bf 186} 1319-1326

  \noindent
  \hangindent=3em
  Novikov I D 1974 Isotropization of homogeneous cosmological models
  in {\em ``Confrontation of cosmological theories with observational 
    data''}, IAU Symposium No. 63, edited by Longair M S
  (Reidel:Dordrecht) 

  \noindent
  \hangindent=3em
  Wainwright J, Hancock M J and Uggla C 1999 Asymptotic
  self-similarity breaking at late times in cosmology \CQG {\bf 16}
  2577-2598 

  \noindent
  \hangindent=3em
  Wainwright J and Ellis G F R 1997 {\it Dynamical systems in
    cosmology} (Cambridge University Press: Cambridge)

\end{document}